\documentstyle[11pt,tables]{article}
\input psfig
\begin{document}
\begin{center}
{\large \bf Branching Ratios in Proton Antiproton Annihilation
at Rest \\from Large $ \mbox{\boldmath{$ N_c$}}$ QCD } \\
\vspace{1.0cm}
{\bf Yang Lu}  and {\bf R. D. Amado} \\
Department of Physics,  \\
University of Pennsylvania, Philadelphia, PA 19104, USA  \\
\vspace{0.3cm}
\today
\end{center}
\begin{abstract}
We use classical or large $N_c$ QCD to describe the mesons 
($\pi$, $\rho$, $\omega$) coming from proton-antiproton
annihilation at rest as classical fields, which we then quantize as 
coherent states.  This treatment gives a nearly parameter
free account of the pion branching  ratios in annihilation.  
\end{abstract}  
\section{Introduction}
Proton antiproton annihilation at rest goes through many
channels, yielding mostly pions.
Some 33\% of these 
pions are the secondary products
of annihilation first into meson resonances, principally rho and omega.    
The total number of pions is large. Kinematically, 
it can vary from two to thirteen, but
the average number is about five with a variance of one. 
A challenge to theory is to calculate the pion spectrum 
and the branching ratios of the many annihilation channels.  
This calculation should be done in the context of the fundamental
theory of the strong interactions, QCD, 
noting that annihilation at rest is 
squarely in the domain of nonperturbative QCD. 
In this note we exploit the large pion number to approximate the pion
field as classical.  This leads naturally to classical QCD, which
is equivalent to QCD in the  large $N_C$ 
(number of colors) limit, the appropriate limit for
nonperturbative QCD \cite{'tHooft,Witten}. We generate the pion,
rho and omega classical fields from annihilation dynamically  and
then quantize the fields in the asymptotic or free field region  
using the method of coherent states generalized to respect
isospin and four momentum conservation. From this state we obtain 
predictions about the branching ratios, momentum spectrum,
meson number and charge type distributions.  This calculation 
is natural in our unified treatment
since we have only one state that describes the annihilation, and
all the different channels emerge from projections onto that state.  
We find remarkable agreement with the data, in particular we correctly
reproduce the trends in the branching
ratios to the many annihilation channels. The
only parameter we adjust is the size of the annihilation
region, and that comes out at a reasonable value. 

We have previously shown how large $N_C$ QCD can be used to 
describe  annihilation  first with pions only in the context of
the Skyrme model \cite{Skyrme,ACDLS} and then how this 
picture can be extended to include the $\omega$ meson \cite{bs&rda}.
Here we extend this work to include the $\rho$ meson 
as well \cite{detail} but more importantly we make an effort 
to connect with experiment, in particular the branching ratios 
\cite{exp,Amsler,Dover}. 
Our previous work explored the theoretical
approach of classical QCD for the dynamics, followed by quantum
coherent states to impose quantum numbers and to obtain field
quanta. This note establishes that approach as sensible phenomenology.

It is well known that large $N_C$ QCD is a classical field 
theory of mesons only, in which baryons emerge as topologically 
stable nonperturbative configurations of the meson 
field \cite{'tHooft,Witten}. The Skyrme model \cite{Skyrme}
is the best known example of a theory incorporating
these features with pions only.  It begins with the 
non-linear sigma model, a general feature of all classical 
QCD pictures, and adds a fourth derivative term, called the 
Skyrme term, to stabilize the baryons.  This added term is,
probably, the first in a series of terms with more and 
more derivatives the exact nature of which  
is not known since no one has yet derived classical QCD from the
quantum theory.  Nevertheless, it is generally agreed 
that the features of any such theory at low energies will 
be those of the Skyrme like models, the higher derivative 
terms only affecting the details at higher momenta. 
Annihilation at rest is, in this sense, a low energy phenomenon
and we believe we can therefore be led by the first few terms. 
It is this robust nature of Skyrme like models at low 
momentum that we exploit in this note.
To include the $\rho$ and $\omega$ classical fields in the large $N_C$
treatment, we treat them as massive Yang-Mills fields which gauge the 
$U_V(2)$ symmetry of the non-linear $\sigma$ model \cite{BZ,MK}. 
The couplings between
the vector meson fields and the pion field are fixed by the KSFR relation,
\cite{KSFR}, in terms of the vector meson mass (we take $m_{\rho} =
m_{\omega} = 770$ MeV) and $f_{\pi}$, the pion decay constant. 
We follow the usual convention in Skyrme calculations and
fix this constant at $f_{\pi}= 75$ MeV to give the 
nucleon mass \cite{BZ}. The vector mesons stabilize the baryons, thus
eliminating the need for the Skyrme term. Hence our dynamics is completely
specified by only three parameters, the pion mass, the nucleon mass and
the vector meson mass, all three of which we fix at their observed values.  

Studies of annihilation in the Skyrme model have shown that when a Skyrmion
and antiSkyrmion annihilate \cite{SSLK}, or when a ``blob" 
of Skyrmionic matter with zero
baryon number but with energy and size appropriate to a
nucleon-antinucleon pair evolves
\cite{SWA}, pion
radiation emerges as quickly as causality will permit.  This picture of
annihilation as proceeding in a coherent burst of classical radiation is the
opposite of the traditional thermal fireball decay picture.  We shall see
that constructing a quantum coherent state based 
on the rapid burst of radiation
from the classical dynamics gives an excellent account of annihilation 
channels.  We construct this account in a simplified 
picture in which we begin with a spherically symmetric blob of 
pionic matter with total energy of two nucleon masses, and at rest.  
We use the dynamical equations of classical QCD to evolve the pion 
field and to develop the coupled omega and rho fields. From these 
fields in the radiation zone, we construct our quantum coherent 
state, projected onto good isospin and four momentum.  From this
state we project the various decay channels and find the branching ratios.
By starting with a blob, we skip over the difficult question of 
modeling the annihilation process itself and computing rates.  
We also cannot study Bose-Einstein correlations in this simplified 
picture \cite{ACDLL}. Some compensation for these shortcomings 
can be found in that the blob starting point introduces only 
one new parameter, the size of the blob.  We use this as an adjustable
parameter and fit it to the pion momentum spectrum.  
The 
branching rates are then calculated
with the parameters of the theory fixed by the three masses 
(pion, nucleon and vector meson) and by the size of the 
annihilation region, determined through the one pion spectrum. 

\section{Calculation}
To begin our dynamical calculations we need an initial pion field
configuration.
As before \cite{bs&rda} we parameterize that configuration as
\begin{equation}
F(r,t=0) = h \frac{r}{r^2+a^2} \exp(-r/a)
\end{equation}
where $a$ is a range parameter and $h$ is fixed by the 
condition that the total energy be twice the proton rest mass.  
This form for $F$  guarantees that the initial baryon number is zero.  
Note that the initial $\rho$ and $\omega$
fields are chosen to be zero.  The $\rho$'s and $\omega$'s seen in
annihilation are generated dynamically by 
their interaction with the evolving
pion field. The final pion momentum distribution reflects our
choice of $F$, but is by no means identical to it.  Rather that
final distribution has $F$ passed through the non-linear dynamical
equations.
Our previous work was directed more at formal exposition than
phenomenology, and we chose $a=1/m_{\pi}$. We
found that this choice of $a$ led to a mean pion number
closer to 7 than the experimental 5 and to a one body
pion distribution in momentum space that
peaks somewhat below the experimental value.  
These two discrepancies are correlated and 
it is clear that a smaller $a$ will
lead to higher mean momentum for the pions 
and correspondingly smaller
average number of pions.  We have carried out the 
dynamical calculation \cite{detail} for
a range of smaller $a$'s and compared the 
resulting asymptotic pion distributions with experiment.  
We find that $a=1.1$ fm gives a very good fit to the 
experimental momentum distribution as seen in Figure 1.  
It also leads to a mean pion number of 5 with variance 1, 
both in close agreement with experiment. We should note 
in passing that our simple one parameter
form for the initial pion field is good enough to account for the
observed pion momentum distribution.  A range of $1.1$ fm
is also a more reasonable size for  the
annihilation region than a pion Compton wave length. Having fixed the
size of the annihilation region by the pion momentum distribution, there
are {\bf no free parameters} left in our calculation.  The
other parameters are fixed by the masses.

Given the initial pion configuration we use the classical dynamical
equations to generate the 
asymptotic  $\pi$, $\rho$ and $\omega$ fields.
From them we  construct
the quantum coherent state corresponding to those fields. 
We use the projection techniques we have 
exploited before to define states of good four-momentum and 
isospin \cite{ACDLS,HornSilver,BSS}. Since it is the
total isospin we wish to specify (recall that the 
$\bar{p} p$ system can have either $I=0$ or $I=1$), and since 
both the $\rho$ field and $\pi$ field carry isospin, we need to 
project each of those fields and then combine the projected states 
using standard Clebsch-Gordan techniques.  From all this comes
a state $| I,I_z,I_1,I_2;K>$ where $I$ and $I_z$ are the total i-spin and
z-component (for $\bar{p} p$ we have $I_z =0$) and $I_1$ and $I_2$ are the
total i-spins carried by the $\pi$ and $\rho$ states. In terms of these states
we can calculate the amplitude for finding a state of a fixed number of
$\pi$ mesons of each charge type and fixed momentum, a fixed number of
$\rho$ mesons of given charge type and momentum and a fixed number of
$\omega$ mesons of given momentum.  This amplitude is just the overlap of
a state of the specified meson number, type, charge and momentum with
the projected (normalized) coherent state.  The probability, then, of finding
a fixed number of $\pi$ mesons of each charge type, a fixed number of $\rho$
mesons of specified charge type and a stated number of $\omega $ mesons 
in a state of fixed total i-spin and z-component is
the absolute square of the overlap amplitude integrated over all meson
momenta and summed over all possible values of intermediate $\pi$ and
$\rho$ isospin, $I_1$ and $I_2$.  To evaluate the overlaps and sums while
implementing four-momentum conservation we found it necessary to use
the expansion techniques developed in \cite{ACDLS} and \cite{bs&rda},
extended to the case of three meson types. 

From the calculation outlined above emerge the branching probabilities
for proton antiproton annihilation at rest into each possible set of
$\pi$, $\rho$ and $\omega$ mesons separated by charge type and number.  In
most data compilations, only total pion numbers are quoted.  It is 
straight forward to convert our branching ratios into the three meson types
into pions only noting that each of the vector mesons has a 
principal decay mode into pions: 
$\rho^{\pm} \rightarrow \pi^{\pm} + \pi^0$,
$\rho^0 \rightarrow \pi^+ + \pi^-$, $\omega \rightarrow \pi^+ + \pi^0 +
\pi^-$.  Armed with this formalism, we can compare our results with data.

\section{Results}

Let us look at the pion branching ratios.  They are given in the literature
as percentages.  For the simpler decay modes the branching ratio is for
one particular pion configuration, but decays involving many $\pi^0$s
are lumped together.  In Table 1 we show the 
experimental branching percentages
compared with our calculations.  To compare with the data we need to make
an assumption about isopin. We make the simple assumption of
equal amounts of $I=0$ and $I=1$. Our results would change
very little if we took the mix of 63\% $I=0$ and
37\% $I=1$  suggested by \cite{Milan}.
The fit to the data in Table 1 is striking.  The small channels
come out small and for the large branching ratios, we are in fair, sometimes
excellent 
agreement with experiment.  In fact given the simple and parameter free
nature of our treatment the agreement is surprisingly good. 
Our most serious relative discrepancies are for the channels, 
$\pi^+\pi^-$ and $\pi^+\pi^-\pi^0$. These are both two body channels
since $\pi^+\pi^-\pi^0$ is dominated by $\pi\rho$. We would expect
quantum corrections to our classical treatment to be largest for these
channels involving the fewest quanta.

Also shown in the table is the percentage of secondary pions, that is the
percentage of pions coming from resonance decays.  
These are nearly all from $\rho$ and $\omega$ decay.  The experimental number
is 33\% and we find about 30\%, again in very reasonable agreement.  Without
the dynamical generation of vector mesons,  we could not make contact with
this result.

\section{Conclusions}

We have seen that a dynamical picture based on classical or large $N_C$ QCD
gives a remarkably accurate account of the branching ratios 
to the various pion and vector meson channels in $\bar{p} p$
annihilation at rest.  It does this in the context of a very simple starting
assumption and only one parameter.  The principal new feature
of our treatment is that all annihilation channels are described
in terms of one single coherent state.  This unified view is essential 
in setting the relative scale of the channels.  It might be argued that our 
agreement with experiment simply reflects phase space 
that is implemented by imposing
energy-momentum conservation on the quantum coherent state. 
Phase space is no doubt important particularly in distinguishing the
large from the small channels.  However, it is by no means 
the whole story since the vector meson fields are developed dynamically in
our picture.  Without those vector mesons we would not come close
to fitting the data.  We have done a ``pions only" calculation fit to
the pion spectrum, and find a far less good description of 
the channels \cite{detail}.
Thus some aspects of the QCD dynamics are essential to our picture.

What we have done is certainly only the beginning of a complete treatment
based on the approach of solving the difficult dynamics using classical
QCD and quantizing afterwards to make contact with experiment. Much more
remains to be done in applying the full power of this method to annihilation.
For example we need to work to obtain detailed agreement with the 
branching ratios. This would require better treatment of quantum corrections. 
We should also study annihilation in flight and take account the 
two center nature
of the process both to obtain rates and to study 
Bose-Einstein correlations \cite{ACDLL}.
To account for the channels with strange particles 
(about 7\% experimentally),
we need to consider large $N_c$ QCD with $SU(3)$ flavor symmetry.
We are beginning such studies.  
Further afield we note the remarkable correlations 
found in annihilation from polarized protons between the proton spin
direction and the charge of pions seen in the rare two pion annihilation 
mode \cite{Hasan}.
Such correlations are difficult to explain in most pictures. However
spatial-isospin correlations are a natural feature of the nonperturbative
baryons generated in large $N_C$ QCD.  We are trying to extract the 
experimentally observed
correlations from our treatment.  Applications of classical QCD 
to processes
involving many pions has been suggested in a number of other contexts,
including the disoriented chiral condensate \cite{DCC} and very high
energy heavy ion collisions.  In this later case, the spin-charge
correlations of classical QCD have also been noted \cite{B&D}.

In summary we have shown that starting from the dynamics of classical QCD
we can construct quantum coherent states that account for the principal
features of proton antiproton annihilation at rest including the 
branching ratios to nearly all channels, large and small.  The
main feature of our treatment is that it treats all these channels in
terms of a single quantum state and that that state is constructed from
dynamical information obtained from classical QCD.  Given the
single pion momentum distribution from annihilation, there are
no free parameters in our treatment.  \\

 We thank Milan Locher for helpful comments and a critical reading 
of the manuscript.  This work is partially supported by the United 
States National Science Foundation.

\newpage

\begin{figure}
\centerline{\hbox{
\psfig{figure=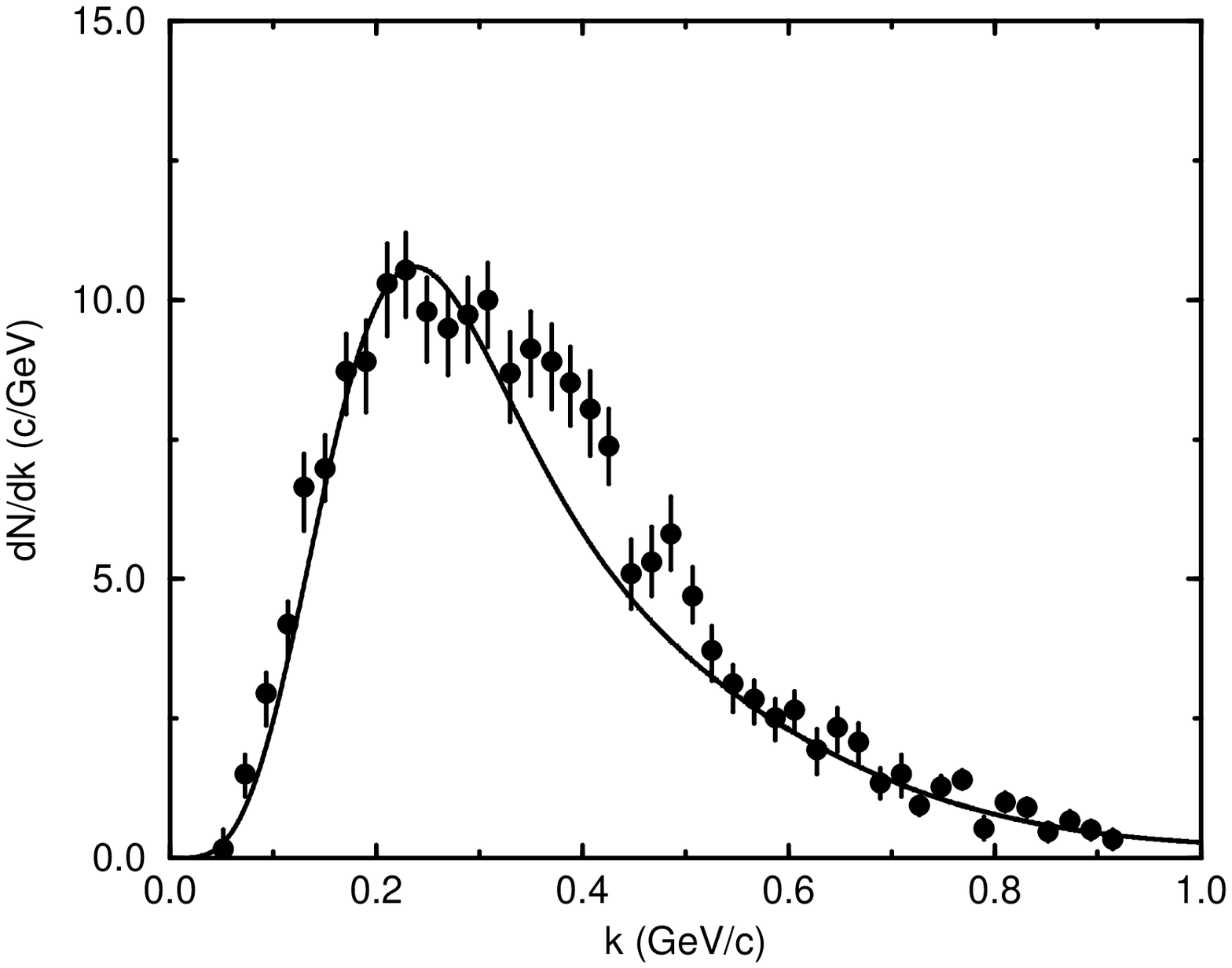,height=2.7in}
}}

\caption{The inclusive single pion momentum spectrum from proton 
antiproton annihilation at rest.  The solid curve is  
our calculation while the data are from \protect{\cite{Dover}}.}
\end{figure}

\begin{table}

\begin{center}
\begin{tabular}{|c|ccc|cc|}   \hline
     & 
\multicolumn{3}{c|}{Theory} & \multicolumn{2}{c|} {Experiment}
\\  
Channel & $I=0$ & $I=1$ & Combined & CERN & BNL \\ \hline
$\pi^+\pi^-$        & 0.02 & 0.0  & 0.01 & 
$0.37\pm0.3$ &$0.32\pm0.04$ \\ \hline
$\pi^+\pi^-\pi^0$   & 0.04 & 0.6  & 0.32 & 
$6.9\pm0.35$ & $7.3\pm0.9  $ \\ \hline 
$2\pi^+2\pi^-$        & 9.1 & 3.0 & 6.1    & 
$6.9\pm0.6$ & $5.8\pm0.3  $ \\ \hline
$2\pi^+2\pi^-\pi^0$ & 26.8 & 19.8 & 23.3 &
$19.6\pm0.7$ & $18.7\pm 0.9$ \\ \hline
$3\pi^+3\pi^-$       & 13.8 & 3.56 & 8.7 &
$2.1\pm0.2$ & $1.9\pm0.2$ \\ \hline
$3\pi^+3\pi^-\pi^0$ &  4.38 & 0.61 & 2.5&
$1.9\pm0.2$ & $1.6\pm0.2$ \\ \hline
${\rm n}\pi^0$, ${\rm n}>1$ & 7.7 & 15.7 & 11.7 &
$4.1\pm0.4$ & $3.3\pm0.2$ \\ \hline
$\pi^+\pi^-{\rm n}\pi^0$, ${\rm n}>1$ & 25.1 & 39.8 & 32.5 &
$ 35.8\pm0.8$ & $34.5\pm1.2$ \\ \hline
$2\pi^+2\pi^-{\rm n}\pi^0$, ${\rm n}>1$ & 12.8 & 17.4 & 15.2 &
$ 20.8\pm0.7$ & $21.3\pm 1.1$ \\ \hline
$3\pi^+3\pi^-{\rm n}\pi^0$, ${\rm n}>1$ & 0.03 & 0.014 & 0.022 &
$0.3\pm0.1$ & $0.3\pm0.1$\\ \hline
\% of secondary $\pi$s & 29.2 & 31.3 & 30.3 
& \multicolumn{2}{c|}{33} \\ \hline
\end{tabular}
\end{center}

\caption{Branching ratios, in percent, for proton antiproton annihilation
at rest.  Our calculations are compared with experiments from 
\protect{\cite{exp}}.
We show each total isospin channel calculated separately. The ``combined"
column corresponds to equal mixture of $I=0$ and $I=1$.  In the last
row we list the percentage of pions from the decay of rho and omega mesons.}
\end{table}
\end{document}